\documentclass[manuscript, nonacm]{acmart}
\usepackage{url}
\setcopyright{none}  
\acmConference[arXiv Preprint]{arXiv}{July 2025}{}
\acmYear{2025}
\acmDOI{}  
\acmISBN{}  

\title{Toward Neurodivergent-Aware Productivity: A Systems and AI-Based Human-in-the-Loop Framework for ADHD-Affected Professionals}

\author{Raghavendra Deshmukh}
\affiliation{
  \institution{PES University}
  \city{Bangalore}
  \country{India}
}
\email{raghavendradeshmukh@pes.edu}

\begin{document}
\begin{acks}
This work is licensed under a Creative Commons Attribution 4.0 International License (CC BY 4.0). To view a copy of this license, visit \url{https://creativecommons.org/licenses/by/4.0/}.
\end{acks}
\newline

\begin{abstract}
Digital work environments in IT and knowledge-based sectors demand high levels of attention management, task juggling and self-regulation. For adults with Attention-Deficit/Hyperactivity Disorder (ADHD), these settings often amplify existing challenges such as time blindness, digital distraction, emotional reactivity and executive dysfunction. Such individuals prefer a low-touch, easy-to-use interventions to help them in their day-to-day activities. Conventional productivity tools fall short of supporting the cognitive variability and overload patterns experienced by neurodivergent professionals.

This paper introduces a comprehensive framework that blends Systems Thinking, Human-in-the-Loop with Artificial Intelligence (AI), Machine Learning (ML) and privacy-first adaptive agents to support ADHD-affected users in managing digital work. At the heart of the solution is a voice-enabled productivity assistant that senses user behavior—tab usage, application focus, inactivity windows—using lightweight, on-device machine learning. These behavioral cues are analyzed in real-time to infer attention states and deliver adaptive nudges, reflective queries or  accountability-based presence (body doubling) designed to co-regulate cognition without disruption.

While technically grounded in AI/ML, the design is deeply influenced by Systems Thinking, viewing user attention as a product of dynamic feedback loops. This hybrid approach bridges behavioral sensing with cognitive inclusivity, offering a replicable model for adaptive, neurodivergent first decision support tools in high-distraction work environments.

\end{abstract}


\maketitle

\section{Introduction}
It is estimated that 4\% to 5\% of adults globally are affected by ADHD \cite{nimh_adhd}. Among working professionals, the number is likely underreported due to factors such as late diagnosis, masking behaviors, and ongoing stigma in corporate environments. With the rise of remote-first and digital-first work, the challenges associated with ADHD—once partially buffered by in-person environmental scaffolds—have become more visible, disruptive, and harder to manage.
Digital work environments demand sustained attention, executive control, and rapid context switching, conditions that are inherently misaligned with the core difficulties experienced by ADHD-affected individuals. These include time blindness, emotional reactivity, disorganized thinking, and cognitive fatigue. Studies show that executive dysfunction—difficulty with prioritizing, organizing, and initiating tasks—impairs day-to-day workplace functioning \cite{Townsend, amador-campos_executive_2023}, often resulting in missed deadlines, impulsive switching, or difficulty following through.
The fragmented nature of knowledge work, which requires juggling between multiple tools and platforms, contributes to heightened cognitive load and task inertia \cite{adler_structure_2017}. Additionally, many professionals with ADHD experience emotional dysregulation, leading to difficulties managing feedback, peer interactions, and stress in high-performance environments \cite{shaw_emotion_2014}.
These barriers are further amplified by the rise of “hustle culture”—a mode of work that glorifies long hours, constant output, and perpetual availability. Particularly prevalent in tech startups and high-velocity digital teams, this ethos incentivizes overcommitment and hyper-productivity, which can be cognitively unsustainable for neurodivergent workers. A recent PLOS ONE study identified how non-attachmental and hustle-driven work cultures can undermine psychological stability and exacerbate attention-related distress, especially among workers with low emotional buffer capacity \cite{rashid_blessing_2024}. Similar findings from organizational studies point to hustle culture as a path to burnout, highlighting how its pressures erode attention regulation and reduce sustained focus in cognitively diverse teams \cite{zhykovska_hustle_2023}.
Most existing tools for ADHD-affected professionals are structured around task reminders, habit tracking, or productivity logs. Although many incorporate machine learning to detect usage patterns, they often rely on shallow context-awareness and fail to support adaptive emotional scaffolding or dynamic cognitive states \cite{bhandarkar_productivity_2024}. These tools tend to prioritize behavior correction over self-awareness and flexibility.
What is missing is a holistic approach that treats ADHD not as a task deficit but as a systems-level attentional modulation challenge. Current interventions rarely take into account the feedback loops between cognition, emotion, and context. Moreover, most tools lack privacy-first architectures and operate with opaque data collection, a problematic pattern given the vulnerability and stigma around neurodivergence in workplace cultures.
In response, this paper introduces a conceptual framework for an AI-powered, voice-enabled assistant designed for ADHD-affected professionals working in digital settings. The system operates fully on-device and uses lightweight behavioral sensing—such as tab churn, inactivity, and app-switching patterns—to infer attention states and deliver adaptive, soft-touch nudges. These may take the form of reflective prompts, brief accountability cues, or body doubling-inspired co-working signals, all of which are governed by user-controlled settings and local privacy rules.
At its core, the assistant applies systems thinking—viewing attention as an emergent property of interaction, rather than a static user trait. A human-in-the-loop feedback layer allows the tool to evolve with the user’s preferences, maintaining relevance and personal alignment. This hybrid model prioritizes presence, flexibility, and neuroinclusive design, offering an emotionally safe alternative to conventional task-centric interventions.
Finally, to ground the tool in real-world insight, we employed a co-design methodology. A preliminary survey with 25 ADHD-affected professionals revealed key struggles with distraction, motivation, and digital fatigue. Their inputs directly shaped the tool’s behavior logic and user interface. One of the authors—an ADHD-affected software developer—provided continuous “lived input,” anchoring the assistant’s design in contextual authenticity.

\section{Literature Review}
\subsection{Using AI/ML in ADHD support: Trends, diagnosis vs assistive models}
There are many applications to support ADHD which have been built using AI and ML technologies and algorithms.  They have been focussed on diagnosis, classification, symptom prediction.  For example, studies such as those by Chen et al. (2023) and Maniruzzaman et al. (2022) \cite{chen_deep_2019, maniruzzaman_predicting_nodate} have used deep learning to distinguish ADHD vs non-ADHD individuals using clinical, behavioral or  sensor data. These systems are typically built around pathology detection using large labeled datasets in controlled environments.

However, few systems apply AI to everyday support for adult ADHD populations. Some emerging tools like EndeavorRx \cite{noauthor_endeavorrx_nodate} for children or  productivity-focused apps like the one quoted by Bhandarkar et al. (2024) \cite{IEEE10910758} offer reminders or nudges, but remain behavior focused and lack real time adaptation. Moreover, these tools rarely account for cognitive state fluctuation, emotional reactivity or individual rhythms. There exist mobile apps that allow ADHD affected students and adults to set their regular routines and assist them manage their schedules via alerts and insights \cite{knouse_usability_2022, kyriakaki_mobile_2023}.

The proposed work addresses this gap by applying lightweight, privacy-first AI not for adaptive support in professional work environments and not diagnosis. The system proposed here responds to moment-by-moment behavioral patterns such as tab usage or inactivity while preserving user autonomy and data privacy.

\subsection{Productivity Tools for Neurodivergent Users – Challenges with One-Size-Fits-All Design}
A wide range of productivity tools such as Trello, Todoist, Notion, Calendars etc are widely adopted in digital workspaces to help users manage tasks, deadlines and goals. While these systems offer powerful scheduling and collaboration capabilities, they often operate on the assumption of sequential planning, sustained attention and consistent executive functioning.

However, for neurodivergent users, especially those with ADHD, such tools can become sources of overwhelm rather than support. Users report “task paralysis”, fatigue and prioritization confusion when interacting with long, static task boards or nested checklist hierarchies \cite{dicks_adhd_2024}. Moreover, conventional productivity apps offer limited adaptability in response to fluctuating cognitive states, resulting in nudges that are ignored or trigger stress.  As Perry et al. (2024) note, \cite{perry_ai_2024} the challenge lies in designing systems that account for executive load, emotional regulation and behavioral variability. Few mainstream platforms currently offer neuroinclusive personalization or context-sensitive support. 

This paper responds to this gap by proposing a lightweight, adaptive assistant that tailors its interventions based on behavioral context, rather than expecting the user to conform to a fixed workflow model.

\subsection{Co-Design and Human-In-the-Loop in Cognitive Technologies}
Human-in-the-loop (HITL) systems have long been used in critical decision-making domains like aviation, medicine and AI ethics. In recent years, their application has extended into mental health and human-computer interaction, where co-regulation between a user and an intelligent system can enhance agency, reduce cognitive burden and promote sustained engagement \cite{perry_ai_2024, ekellem2024enhancing}.

Within ADHD support tools, however, few designs incorporate real-time user feedback or learning mechanisms beyond initial customization. Most interventions remain static once deployed, lacking the ability to adapt based on evolving attention patterns or task strategies. This stands in contrast to participatory design methodologies emerging from the cognitive accessibility and inclusive UX communities, which emphasize designing with, not just for, neurodivergent users \cite{heublein_designing_2023}. It is also noticed that while AI based systems are considered effective to offer interventions, people prefer a HITL approach \cite{lee_artificial_2025}.

Recent research has emphasized the role of AI-enabled body doubling in supporting task initiation and emotional regulation for ADHD-affected individuals. A design-led study by Malmö University \cite{eugenia_leveraging_2024} developed a chatbot prototype to simulate body doubling, particularly for users with social anxiety or limited access to real-time co-working partners, highlighting the importance of presence-based support, non-judgmental interaction and emotional safety as pillars of neuroinclusive design.

Based by these findings and our own co-design process, our proposed system integrates a human-in-the-loop architecture that includes an optional digital body doubling mode, designed to provide gentle presence, reflective cues and accountability-based support. This dynamic layer allows the assistant to evolve through continuous interaction and user feedback, ensuring it remains context-aware, respectful and emotionally attuned, rather than prescriptive or interruptive.

\subsection{Systems Thinking and Feedback Models in Attention Regulation}
While AI based productivity tools often focus on distinct behaviors (e.g., time-on-task, clickstream patterns) they typically lack a systems-level understanding of how attention emerges and fluctuates over time. Systems thinking offers a valuable lens for modeling attention not as a fixed trait but as an output of interacting feedback loops between internal states, external tools and environmental stimuli. \cite{meadows_thinking_2008, capra_systems_2014}.

In the context of ADHD, attention regulation is profoundly shaped by dynamic interactions—task urgency, emotional context, ambient interruptions and internal motivators. Yet few systems have been designed to observe and respond to these feedback patterns in real time. Models from affective computing \cite{picard_affective_2003} and metacognitive scaffolding in education \cite{greene_theoretical_2007} offer promising analogues but remain underutilized in ADHD related tool design.

This paper introduces a novel approach that treats user attention as a state in flux, responding to behavioral, contextual and reflective triggers. Through real-time sensing and adaptive prompting, the system is designed not as a linear tracker but as a feedback-sensitive assistant inspired by the loop-driven logic of systems thinking. This perspective enables interventions that are timely, non-intrusive and situated within the user's cognitive ecosystem.

\section{Methodology}
The development of this ADHD-focused productivity assistant is based on a multi-dimensional approach that aligns with the central hypothesis: that attention challenges faced by ADHD-affected professionals in digital work environments can be mitigated through adaptive, human-centered and systems-aware interventions. To explore this hypothesis, the research combines theoretical frameworks—systems thinking, human-in-the-loop design and privacy-first machine learning—with a participatory co-design process involving ADHD-affected professionals. This section outlines the guiding design philosophy, the survey methodology and user insight process, the ethical foundations of privacy-preserving AI and the principles behind soft-touch nudging strategies.

\subsection{Framework: Systems Thinking, Human-In-The-Loop and Machine Learning}
The design of the proposed productivity assistant is grounded in a hybrid methodological framework that draws from three key disciplines: systems thinking, human-in-the-loop (HITL) design and lightweight, privacy-preserving machine learning. These perspectives collectively shape not just the functionality of the tool, but its underlying logic, interaction model and user-first philosophy.
Systems thinking offers a foundational lens through which attention is conceptualized not as a binary or static trait, but as a dynamic state influenced by a feedback-rich ecosystem involving tasks, tools, context, emotion and self-regulation. Inspired by Capra and Luisi's (2014) \cite{capra_systems_2014} view of cognition as an emergent, interconnected process and Meadows’ (2008) \cite{meadows_thinking_2008} model of systemic feedback loops, the assistant is designed to recognize and respond to fluctuations in attentional state—such as disengagement, hyperfocus or task-switching inertia—rather than treat these as isolated behaviors. This approach enables the tool to intervene not by enforcing external order, but by helping restore self-regulatory balance within the user's workflow ecosystem.

At the interaction level, the tool is shaped by Human-in-the-Loop (HITL) principles drawn from cognitive technology design and inclusive UX research (Perry et al., 2024; Ekellem, 2025) \cite{perry_ai_2024, ekellem2024enhancing}. Rather than acting as a prescriptive, fully autonomous system, the assistant is built to learn with the user, adapting over time through reflective prompts, soft feedback loops and optional input. The HITL model allows for iterative and adaptive learning and mutual regulation where the assistant adapts to individual routines, preferences and fluctuations in energy or focus, while ensuring the user retains control and customization \cite{tejasvi_smart_2024}.

To operationalize this adaptive logic, the system employs lightweight, on-device machine learning models capable of tracking non-invasive behavioral signals such as tab switching frequency, idle time and application usage. Importantly, the ML component is designed with a privacy-first philosophy: all inference and behavioral modeling occur locally, without transmitting personal data to cloud servers. This approach ensures user autonomy and trust, while enabling the system to build a real-time profile of attention dynamics without relying on intrusive data capture.

Together, this framework enables the development of a cognitive support tool that is not just technically adaptive but ethically aligned, context-aware and deeply rooted in the lived experiences and cognitive rhythms of ADHD-affected professionals.

\subsection{Survey and Co-Design}
To determine the assistant’s design in real-world experiences, a preliminary survey was conducted with about 25 self and professionally diagnosed ADHD-affected individuals, including early-career professionals and mid-senior knowledge workers across technology, software design and engineering domains. The survey included both structured and open-ended questions and focused on themes such as attention management, executive dysfunction, digital work overload, coping mechanisms and expectations from supportive tools.

Responses revealed recurring challenges with task prioritization, context switching fatigue and “attention crash” moments particularly during unstructured work periods and transitions between tasks or meetings. Many participants described digital productivity tools (e.g., calendars, to-do lists, jira, obsidian, pomodoro timers) as either limiting or too passive to be effective, citing a desire for non-judgmental, presence-based support rather than gamified motivation or hard deadlines. Several participants expressed a preference for interventions that are quiet, customizable and responsive to their own behavioral rhythms, rather than time-based triggers. Moreover, participants wanted to have a weekly summary of their usage patterns and wanted quiet check-ins from the system.

In addition to the survey, the co-design process was further informed by the lived experience of the authors, a senior software professional diagnosed with ADHD. Their perspective helped translate user pain points into nuanced design choices—such as the inclusion of “body doubling” mechanics, the avoidance of intrusive notifications and the preference for soft, voice-based engagement over text-based interactions. These participatory elements formed the empathic backbone of the assistant’s design, ensuring that the system reflects neurodivergent realities rather than neurotypical assumptions.

\subsection{Privacy by Design: On device ML, User Sovereignty}

Neurodivergent individuals often experience heightened vulnerability in digital settings due to challenges with attention, impulsivity and sensory integration. As Jones et al. (2023) \cite{jones_privacy_2023} emphasize, privacy breaches and behavioral data misuse can exacerbate stigma and isolation, particularly for users with ADHD or ASD in immersive work environments.

A foundational principle guiding the development of this assistant is privacy by design—an ethical and technical commitment to ensuring that user data is never collected, stored or  transmitted without consent. Given the cognitive vulnerability and masking behaviors often experienced by ADHD-affected professionals in digital settings, the assistant is intentionally designed to operate entirely on-device, without sending behavioral or usage data to external servers or cloud-based inference engines.

Behavioral cues such as application usage, tab switching and idle time are processed using lightweight machine learning models embedded locally within the user’s environment. These models detect attentional state changes without requiring high-volume data transfer or persistent surveillance. No personally identifiable information (PII), browsing content or communication data is analyzed or stored.

This architectural choice is particularly critical given the potential misuse or misinterpretation of behavioral data in workplace contexts. ADHD-affected professionals often experience pressure to mask symptoms such as inattention, hyperfocus or  time blindness traits that could be stigmatized if exposed or misunderstood by employers or colleagues. Storing such sensitive data on cloud environments brings in integrity and reputational challenges that may deter usage altogether.

As part of a systems-informed design, the assistant addresses not only technical privacy but also cognitive and emotional safety, as advocated in neurodiversity centered design literature \cite{jones_privacy_2023}. Users must feel safe enough to engage with the tool as a supportive partner rather than an intrusive monitor.  To build trust and ensure autonomy, users are granted full control over sensing and feedback functions. The assistant can be paused, customized or disabled entirely ensuring that support is always consensual, transparent and self-directed. This approach aligns with the privacy values expressed by survey participants and reflects a commitment to neuroinclusive design ethics. To build trust and ensure autonomy, users are granted full control over sensing and feedback functions. The assistant can be paused, customized or  disabled entirely ensuring that support is always consensual, transparent and self-directed. 

In addition, any behavioral data retained for on-device processing is stored in a non human readable format and may be optionally encrypted at the file system level, further reducing the risk of unintended access or misinterpretation. To reinforce autonomy and safety, users have the option to purge all data stored at any time. This includes activity history, inferences created, nudges etc.  The action of purging is non-reversible and does not impact how the assistant functions.  This reinforces the tool’s alignment with data minimization, user sovereignty and neuroinclusive design ethics.

While prior explorations into AI-supported body doubling \cite{eugenia_leveraging_2024} have shown promise in fostering motivation and task engagement, they often lack explicit strategies for safeguarding user privacy. In contrast, our assistant processes presence cues and behavioral signals entirely on-device, ensuring that emotionally sensitive data—such as attention drift or inactivity—remains local and encrypted. This aligns with ADHD users’ stated preferences for privacy, autonomy and control.

\section{Proposed Framework: A privacy-first and neuro-inclusive productivity assistant}
Based on the methodology of systems-oriented design and participatory insights discussed in the methodology, this section presents the core architecture of the proposed assistant. The framework integrates lightweight on-device sensing, voice-based interaction and adaptive, context-sensitive feedback mechanisms designed to support ADHD-affected professionals in digital work environments.

Unlike conventional productivity tools that rely on rigid scheduling or performance metrics, the assistant functions as a co-regulatory digital companion responding in real time to fluctuations in attention, emotional tone and engagement patterns. Based on systems thinking and human-in-the-loop design, the assistant’s architecture is modular, private by design and highly configurable.

The following subsections detail the key components of the system, the nudging strategies used to support user engagement and the feedback-loop model that governs behavioral adaptation over time.

\subsection{Assistant Components: Enabling Human Centric and Systems Aware Interaction}
The assistant comprises three interlinked components: behavior sensing, voice interface and an adaptive feedback engine, each designed to support the fluctuating attention rhythms and emotional states of ADHD affected professionals. Based onn systems thinking principles, these modules do not function in isolation. Instead, they act as dynamic subsystems within a real-time feedback loop that senses, interprets and responds to user context in a manner that is gentle, situationally aware and emotionally safe.
The Behavior Sensing Module forms the system’s input layer, passively capturing low-friction behavioral signals such as tab switches, application transitions, inactivity windows and focus loss patterns. One of its unique features is the \textbf{“Where Was I?”} activity recall layer, which maintains a lightweight, local history buffer of the user’s last N actions across the digital workspace (e.g., moving from a project tracker to Slack to a spreadsheet). This history can be surfaced at any point through a floating interface (say a button) that responds to voice or click with prompts like:
\textit{“You were last working on your report, then checked email.  Want to return?”}
The Voice Interface acts as both a communication bridge and a companion layer, providing intuitive access to assistance without intrusive visual popups or auditory overload. Instead of enforcing action, it offers soft invitations to reflect or resume and it adapts to the user’s preferred tone be it affirming, playful or calming.
Finally, the Adaptive Feedback Engine serves as the brain of the system, mapping sensed behaviors to timely, configurable nudges or suggestions. It constantly refines its response strategy through a human-in-the-loop feedback model, ensuring that the assistant adapts not only to behavioral patterns but to what feels right for the user in different contexts.
Collectively, these components embody a co-regulatory system—not one that tells users what to do, but one that stays alongside them, ready to support focus, recall or rest in ways that are neuroinclusive, respectful and responsive.

\subsection{Nudging Strategy: Soft Interventions for Adaptive Co-regulation}
Nudging in this system is designed not as behavioral enforcement but as soft cognitive companionship. Drawing from systems-aware modeling, nudges are triggered only when the assistant senses attention drift, fatigue patterns, or cognitive overload. Rather than rigid timers, the system responds to behavioral thresholds extended tab churn, long focus stretches, or missed task re-entry.
These nudges are delivered through the voice interface as non-intrusive invitations rather than commands. They may appear as brief reflections (\textit{“Want to pick up where you left off?”}), environmental presence cues (\textit{“Still here with you”}), or supportive reminders (\textit{“You’ve been going strong—need a breather?”}). Each nudge is emotionally safe and configurable. Users can tune frequency, mute cues during specific hours, or select their preferred nudge style from hype-based encouragement to calm affirmation.
A core part of the nudging system is the DopBoost feature—a dopamine-aligned micro-intervention designed to re-engage motivation and reset fatigue. If the assistant detects prolonged uninterrupted work or subtle signs of disengagement, it can offer: \textit{“Want a quick DopBoost? I’ve got a Glow Factor or Zen Zest ready.”}  Users can pre-select their preferred DopBoost modes, which may include:
\begin{itemize}
    \item Mood Fuel: a favorite playlist or word puzzle link
    \item Zen Zest: a short breathing guide or mindful stretch reminder
    \item Reward Rush: affirmation followed by a positive reinforcement loop
    \item Focus Ritual: 2-minute intentional reset with light soundscapes or visuals
\end{itemize}

Unlike static Pomodoro systems, DopBoosts are context-aware and gently suggested only when needed. This makes them particularly suited for ADHD-affected individuals who may struggle with rigid break patterns but still need cyclical motivational input.
Ultimately, the nudging layer supports autonomy, not automation. It respects silence, accepts rejection, and adapts over time shaping itself as a co-regulatory peer, not a taskmaster.

\subsection{Digital Body Doubling and Accountability Presence}
Among the assistant’s most distinctive features is its support for digital body doubling a technique inspired by peer-based accountability strategies frequently used in ADHD communities. Traditional body doubling involves the silent presence of another person during tasks, helping individuals sustain focus, initiate action, and resist avoidance. In our system, this concept is translated into a context-aware co-presence experience, designed not to simulate conversation, but to anchor attention through ambient support and rhythmic affirmation.
When enabled, body doubling mode activates a low-friction companion rhythm, offering subtle cues that affirm presence without distraction. These may include periodic voice-based affirmations (\textit{“Still with you let’s keep going”}), soft ambient tones, or brief check-ins triggered after long stretches of inactivity or tab cycling. Users can configure the style, tone, and frequency of these cues to match their preference whether it's a calm guide, an energetic motivator, or a silent rhythmic pulse. This allows the assistant to serve as a scaffold for executive function, helping users transition into tasks, persist through difficulty, and recover focus without judgment.
Theoretically, this feature draws on concepts from social computing, affective presence, and attention regulation. Studies in human-computer interaction suggest that even minimal ambient cues of shared presence can improve task engagement and reduce perceived isolation, especially in remote and asynchronous work settings. For ADHD-affected users, the absence of real-time co-working structures can exacerbate time blindness, motivational inertia and task initiation difficulty. Digital body doubling acts as a counterbalance, offering externalized presence that modulates emotional overwhelm and supports action initiation.
From a systems perspective, body doubling operates as a feedback loop intervention, not a linear prompt. It senses moments of drift, passivity, or cognitive load, and responds with co-regulatory cues that gently re-anchor the user’s attention. Over time, these cues can adapt based on user feedback—some may prefer less frequent affirmations, while others find rhythmic nudges helpful to maintain flow.
This mode is not framed as a timer or focus lock; instead, it is designed as a non-intrusive presence layer that honors user autonomy while supporting attention as a fluctuating, relational state. Future iterations may allow for multi-user body doubling, where two or more neurodivergent users synchronize their assistants to support co-working remotely while maintaining privacy boundaries.

Some of the ways the digital body doubling can interact with the user are as below: 
\begin{itemize}
    \item Light gentle verbal affirmations (e.g., \textit{“Still here with you,” “Let’s keep going”}).
    \item Passive presence cues (e.g., \textit{ambient tones at intervals}).
    \item Gentle accountability prompts (e.g., \textit{“Would you like to reflect on that tab you’ve reopened 3 times?”}).
    \item Voice based reflections if signs of fatigue or drift in attention are detected (e.g., \textit{““You’ve been circling between tasks. Want to reset or re-center?”}).
\end{itemize}

\subsection{Feedback loops: Modelling Cognitive States through Systems Design}
At the core of the assistant’s logic is a systems-inspired feedback model that treats user attention not as a fixed trait, but as an emergent, modifiable state. This state is shaped through continuous interaction between digital tools, task context, emotional regulation, and environmental inputs. Drawing from cybernetics and learning sciences, this approach repositions attention away from compliance-based models (e.g., time-on-task) and toward dynamic state modulation, where the goal is stability, recovery, and adaptation.
The assistant leverages its behavioral sensing layer as a real-time input mechanism, detecting patterns such as tab churn, pause thresholds, re-activation sequences, and prolonged idle states. These indicators are passed through a lightweight inference model that maps them to cognitive state profiles—such as drift, hyperfocus, emotional fatigue or decision inertia. The system then selects a corresponding micro-intervention, such as a reflective prompt, soft accountability cue, or passive ambient nudge. Importantly, the assistant allows users to calibrate the sensitivity and frequency of these looped interventions.
Unlike traditional productivity systems that log tasks or generate scores, the assistant does not accumulate history as a performance ledger. It does not rank, penalize or gamify. Instead, it operates as a real-time, non-evaluative support loop, with each cycle focused on detecting subtle transitions in user engagement and gently steering toward re-regulation.
This design supports a model of “fluid cognitive scaffolding” where interventions are offered when needed, withdrawn when not, and always subject to user override. Over time, the system can learn user-specific patterns: for instance, distinguishing between intentional multitasking and fragmented attention. This adaptive learning supports a broader vision of the assistant as a contextually intelligent companion, helping users build rhythms that align with their neurocognitive patterns, rather than override them.

\subsection{AI and ML Techniques for User Activity Monitoring and Behavior Analysis}
To develop the assistant capable of monitoring and analyzing user behavior on laptops and desktops, several advanced machine learning (ML) and AI techniques can be employed. Given that the user's activity is time-dependent, one critical aspect of the tool is time-series analysis to track behavior patterns over time. Recurrent Neural Networks (RNNs), specifically Long Short-Term Memory (LSTM) networks, are ideal for modeling sequential data, such as switching between apps or tabs, as they can capture dependencies and trends in user behavior over extended periods. LSTMs are well-suited to handle the temporal aspect of the data, making them effective in detecting periods of intense engagement (rapid switching) versus more focused work periods. LSTM-based models have been used successfully in various behavioral sequence detection tasks, including recognizing activity patterns and predicting future behaviors in time-series data. [Ref: Kannagi paper]

To train such models, labeled datasets of user behavior would be required. This dataset could include time-stamped logs of app and tab usage, window switching and web browsing activities. For a robust model, data labeling could be based on predefined categories such as "focus mode," "multitasking," or "distraction" based on specific thresholds of behavior, like the number of windows open or the frequency of switching. Alternatively, user self-reporting or external feedback (e.g., productivity surveys) could be used for labeling to ensure that the model learns to differentiate between productive and non-productive behavior. This labeled data can then be fed into the LSTM model for training, allowing the system to predict and recognize the user’s state at any given time.

Additionally, activity recognition is another essential aspect of the tool, which involves classifying different user behaviors based on usage patterns. Machine learning algorithms like Random Forests, Support Vector Machines (SVMs), and k-Nearest Neighbors (k-NN) can be applied for this classification task. These models are especially useful for differentiating between periods of intense focus versus moments of distraction \cite{mcdonald_classification_2020}. These classifiers can learn from a variety of features, such as the number of apps open, the frequency of window switching, or the length of time spent on a particular task. Feature engineering would play a crucial role here, as relevant features need to be extracted from raw data, such as counts of active windows, app usage time, and engagement duration with individual apps or websites.
For model training, the first step is to collect a sufficiently large dataset of user activities. This can be achieved through user logging applications that track window switching and app usage behavior across different times of the day. Data could be gathered through a lightweight, privacy-preserving background application that logs user behavior locally, ensuring that no sensitive information is transmitted externally. The data would then be annotated (either manually or semi-automatically) to mark periods of focused work, multitasking, or browsing. Once the dataset is prepared, models can be trained using supervised learning techniques, with algorithms like SVMs or Random Forests.
Anomaly detection techniques, such as Isolation Forests, One-Class SVM, and Autoencoders, can be applied to identify abnormal patterns in user behavior. These algorithms are well-suited to flag behaviors that deviate from typical usage, such as excessive multitasking or prolonged engagement with non-work-related activities, which may indicate burnout or decreased productivity. In training anomaly detection models, normal behavioral patterns need to be established through representative datasets, which can be generated by tracking users over a period of time to capture typical engagement patterns \cite{liu_isolation-based_2012}. Outliers such as sudden spikes in browser tab switching or abnormal work patterns—can then be detected as anomalies, prompting the system to recommend a break or intervention.
Another critical component of the system is reinforcement learning (RL), which can be applied to personalize and optimize break suggestions and nudges. By learning from user responses to previous recommendations, the system can tailor the timing, frequency, and content of interventions to maximize effectiveness. This process involves training RL models with reward functions based on the success of previous suggestions (e.g., user productivity after a break) and adjusting the policy over time to provide more contextually relevant reminders. The model would require continuous feedback, which could come from user behavior (e.g., improved focus after a break) or explicit ratings provided by the user on the utility of the break suggestions \cite{hassouni_personalization_2018}. This allows the system to become increasingly efficient at suggesting breaks or productivity tips as it learns the user's preferences and response patterns.
To enhance the tool's context-awareness, the system needs to track the last accessed apps and browser tabs, so it can provide reminders based on the user’s recent activities. A context-aware model might use semantic understanding of the user's recent work, enabling it to provide more accurate nudges. For instance, if a user has been engaging in work-related tasks, a reminder might encourage them to stay focused, while a reminder after a long period of social media browsing might encourage a productive break. Context-aware systems that leverage this kind of information have been widely used in recommendation systems to improve engagement and productivity \cite{canada_item_2022}.

\subsection{Datasets and Model Training}
The development of these models requires access to high-quality, privacy-preserving datasets. For behavior monitoring, datasets should include app usage logs, tab switching frequencies, window activity times, and browsing habits. Collecting these datasets requires ensuring that no personal data is transmitted or stored externally, aligning with the privacy goals of the tool. Additionally, datasets for training anomaly detection and activity recognition algorithms could be augmented with user self-reports, where users provide feedback on their perceived productivity and focus levels, or by employing user surveys that capture how users engage with the tool over time.
For training the models, a combination of supervised learning for activity classification (e.g., focused vs. distracted behavior) and unsupervised learning for anomaly detection would be effective. Pretrained models for similar tasks, such as user activity recognition (e.g., mobile usage datasets for activity recognition), can be adapted for this task by fine-tuning on the laptop/desktop-specific data. Several existing datasets for user activity logging, such as the BID (Behavioral Information and Data) dataset or MULTI (Multitasking Interaction) dataset, could provide foundational data for training models in activity recognition and multitasking behavior.
By using these AI/ML techniques, datasets and privacy-preserving strategies, the tool can provide meaningful, personalized recommendations to users while respecting their privacy. It can dynamically track behavior, detect patterns and make timely interventions that encourage healthier, more productive working habits.

Datasets play a key role in helping the assistant get the right kind of pattern recognition across time for the various activities that the user shall perform.  Task switching and multi tasking research from the perspective of using computers and work-from-home remote meetings yield a good set of informative and interesting datasets \cite{cao_large_2021, zhang_look_2015}.
The ADHD behavior information and data (BID) dataset is another rich source of behavioral data and offers various patterns of user behaviors. The Hyperaktiv by Simula is another rich source of activity data of 100+ people's activity where the source of data is from 51 ADHD patients and 52 people in clinical trials \cite{hicks_hyperaktiv:_2021}. Additionally sources like Kaggle, UCI Machine Learning etc have supplimentary datasets on behaviors, attention switching etc. 

\subsection{System Design}
The high level design component design for the assistant is highlighted in the following diagrams. 

\begin{figure}[ht]
  \centering
  \includegraphics[width=0.8\linewidth]{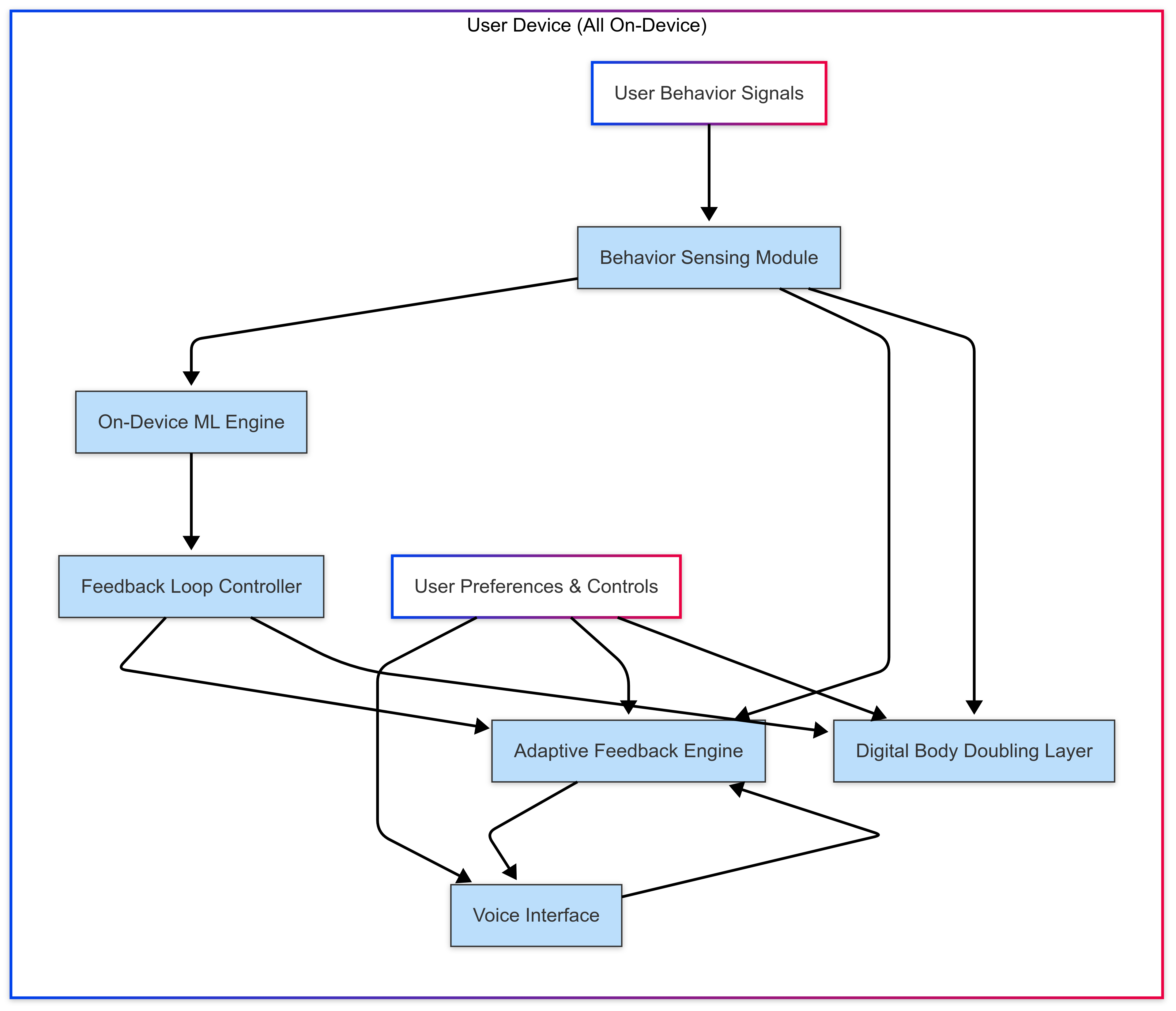}
  \caption{Proposed ADHD productivity assistant framework.}
  \Description{Diagram showing a loop between the user, environment, AI module and nudging system.}
  \label{fig:adhd-framework}
\end{figure}

Figure~\ref{fig:adhd-framework} shows the assistant's components that interact with the user and each other to deliver the on-device experience for the user. The user can control their preferences which are used to control the feedback engine, voice interface and signal the use of digital body double. 

\begin{figure}[ht]
  \centering
  \includegraphics[width=0.8\linewidth]{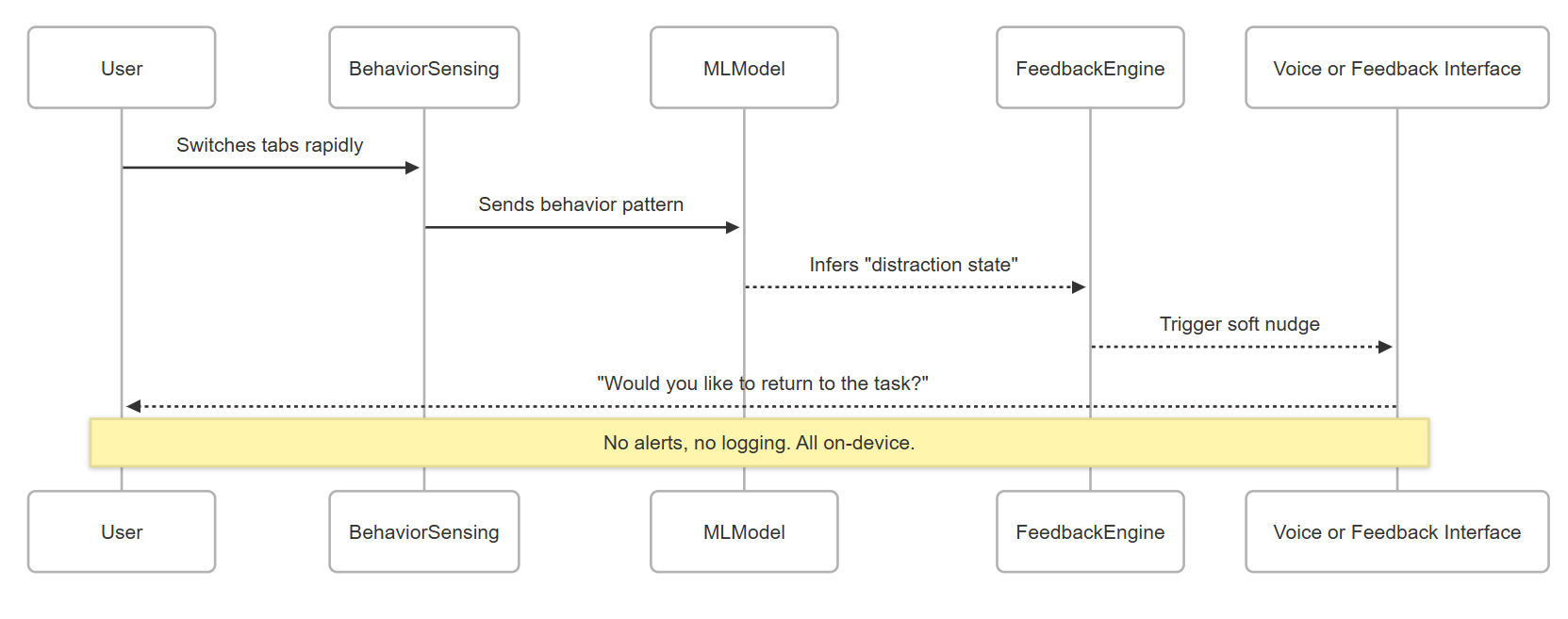}
  \caption{Soft Nudge Sequence Diagram.}
  \Description{The Sequence diagram for the soft nudge that the user receives based on their preferences}
  \label{fig:seq1}
\end{figure}

Figure~\ref{fig:seq1} shows the soft nudge workflow for the users based on their preferences.  Here the behavior sensing module will understand the behavior patterns and uses the on-device ML model to determine a distraction state and use the feedback interface to trigger a soft nudge to the user.

\begin{figure}[ht]
  \centering
  \includegraphics[width=0.8\linewidth]{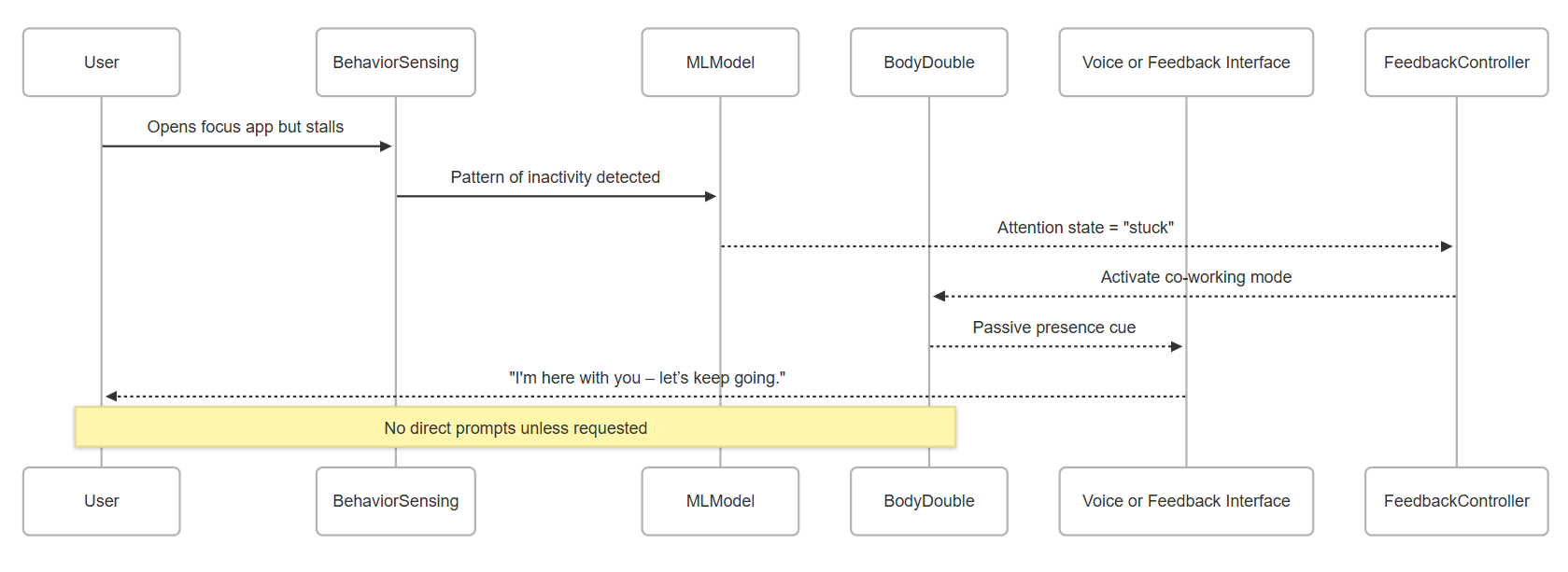}
  \caption{Digital Body Doubling Sequence Diagram.}
  \Description{The Sequence diagram for the digital body double mode being activated based on the user preferences.}
  \label{fig:seq2}
\end{figure}

Figure~\ref{fig:seq2} shows the behavior sensing module using the preferences to understand inactivity pattern and offer a digital body double option.

\section{Survey insights and user informed design decisions}
To anchor the assistant to the everyday experiences of ADHD-affected professionals, a survey was conducted with 25 ADHD-affected digital professionals, primarily from the software engineering and IT services industry. Participants ranged from 21 to 54 years and included both professionally diagnosed and self-diagnosed individuals. Their work contexts included remote, hybrid and in-office roles with digital tools playing a central role in their workflows.

Respondents regularly used devices such as laptops, desktops and smartphones and were highly dependent on multi-tab browsing and application switching. This population represents the real-world demographic most affected by digital work fragmentation and executive dysfunction.

\subsection{Challenges reported in digital work}
The survey revealed recurring cognitive challenges that align closely with existing ADHD research:\newline

\begin{itemize}
    \item Tab overload and bookmarking paralysis: A significant number of users (37\%) reported having 21+ browser tabs open at any given time, with 22\% of them reporting having 12 to 20 browser tabs open at any given time and 20\% saying that they “do not track” their tab usage, \textbf{indicating a lack of awareness or control}. Over 80\% also reported bookmarking content with the intention to return later, but never doing so.
    \item Attention fragmentation and task switching: More than 70\% of respondents admitted to being constantly or frequently pulled away from high-priority tasks owing to distractions or lower-priority work. \textbf{This confirms a core executive dysfunction pattern}.
    \item Tool fatigue and productivity shame: While some users employed productivity apps (e.g., Notes, calendar reminders) several reported that such tools often felt rigid, impersonal or  guilt-inducing. Tools designed for neurotypical workflows were seen as overwhelming or ineffective.
\end{itemize}
A few users commented:
\begin{itemize}
    \item \textit{“I often don't see meeting reminders when I'm switching between apps.”}
    \item \textit{“It’s just so hard to not drift away from videocalls.”}
    \item \textit{“I get 8 hours of work done in 1 or 2 hours or  I get nothing done for days.”}
\end{itemize}

\subsection{Design preferences and feedback expectations}
When asked about an on-device assistant that detects attention states and offers gentle nudges, Most participants responded with “Maybe, depending on how it works,” indicating openness if privacy, customization and tone were appropriate. Preferred nudging modes included:
\begin{itemize}
    \item Gentle pop-up reminders - 55\%
    \item Scheduled quiet check-ins - 54\%
    \item Weekly summaries and usage patterns - 59\%
\end{itemize}
Respondents strongly rejected intrusive automation, emphasizing the need for soft, emotionally safe support rather than performance tracking.
A user remarked as below:\newline
\textit{The main problem I would find with tools is I strongly dislike being told or suggested what to do, even if that thing is the right thing to do. Any tool designed to help me should focus on accountability, not prompting or suggestions. For example, If I tell the tool that I'm going to try to get 2hrs of work done this afternoon and "quiet mode" will help, knowing the tool will check up on me can often give me the drive to do it.}

\subsection{Privacy and trust signals}
The importance of privacy emerged as a clear priority from the respondents:
\begin{itemize}
    \item 41\% said that it is very important and 36\% said that it is mandatory, which makes it a very healthy 77\% who prioritize privacy.  This describes privacy to be a very critical in adopting any cognitive support tool.
    \item Participants expressed discomfort with telemetry based systems that monitor behavior or transmit data externally.
\end{itemize}
\textbf{This validates the assistant’s on-device, privacy-first design and supports the architecture choice that favour local ML inference, user-controlled sensing and non-human-readable data storage.}

\subsection{Supplementary insights from online ADHD communities}
In addition to the survey, reflections from public ADHD-focused Reddit threads (e.g., \texttt{r/adhd\_Programmers}, \texttt{r/ADHD}) were reviewed to deepen qualitative understanding. While not formally analyzed, these anecdotes echoed key themes:
\begin{itemize}
    \item Emotional burnout from overusing “productivity” systems 
    \item The desire for tools that simply “keep me company quietly”
    \item Community-driven acknowledgement for body doubling, silent co-working and soft accountability cues
\end{itemize}
\textbf{These narratives supported the assistant’s inclusion of a digital body doubling mode, reinforcing the need for emotional presence without performance pressure.}

\subsection{Linking Insights to Design Decisions}
\begin{table}[ht]
  \caption{User Insights and Corresponding Design Decisions}
  \label{tab:user-insights}
  \begin{tabular*}{\linewidth}{@{\extracolsep{\fill}} p{0.45\linewidth} p{0.45\linewidth}}
    \toprule
    \textbf{Insight} & \textbf{Design Decision} \\
    \midrule
    Users struggle with task-switching and tab overload &
    Behavioral sensing module detects rapid switching, inactivity \\
    
    Users bookmark without follow-through &
    Nudges suggest revisiting uncompleted tasks contextually \\
    
    Traditional tools feel overwhelming &
    System offers soft-touch voice nudges and co-regulatory presence \\
    
    Privacy is critical &
    All behavior modelling is performed on-device, with no telemetry \\
    
    Users want presence, not pressure &
    Ambient voice affirmations and body doubling simulate companionship \\
    \bottomrule
  \end{tabular*}
\end{table}

\subsection{Survey Takeaways}
The survey confirmed that ADHD-affected digital workers face unique challenges that are not addressed by mainstream productivity tools. Their input directly shaped the assistant’s architecture specifically the use of voice-based interaction, soft adaptive nudging, privacy-first local sensing and presence-based support. These insights bridge the gap between theoretical design goals and real-world user needs making the assistant not just inclusive, but authentically user-aligned.

\section{Discussion and Implications}
As digital work environments continue to evolve, so too must the tools that support cognitive regulation, especially for neurodivergent professionals. This section reflects on the broader implications of our proposed assistant, exploring how it challenges conventional paradigms of productivity, how it may reshape collaborative norms within teams and how its core design principles could generalize to other neurodivergent populations. Through these lenses, we reframe the role of AI in workplace cognition and highlight opportunities for inclusive innovation in attention-aware technologies.

\subsection{Reframing AI/ML as a Co-regulator and not a Task Manager}
A central contribution of this work is its repositioning of AI and machine learning—not as task automation engines, but as co-regulators within a neuroinclusive cognitive ecosystem. Traditional productivity tools powered by AI tend to emphasize efficiency: optimizing calendars, tracking completion rates, and triggering reminders based on surface behaviors. While helpful in some contexts, such tools often overlook the emotional and attentional scaffolding required by ADHD-affected users. By contrast, our assistant deploys lightweight, on-device ML not to enforce structure, but to detect patterns of overload, attention drift, and behavioral rhythms—responding with empathetic, user-aligned support.
This approach draws from the emerging domain of affective computing and relational AI, where agents are designed to understand and respond to human emotional and cognitive states rather than operate in strictly transactional terms. Inspired by the ethos of companion technologies, our system emphasizes presence, adaptability and consent over control, integrating with the user’s flow rather than attempting to dictate it. The assistant’s interventions—whether nudges, DopBoosts or digital body doubling function more like a gentle colleague or rhythm partner than a taskmaster.
Conceptually, this also reflects a broader systems thinking ethos: attention is not a trait to be optimized but a process to be supported. ML in this model becomes a background participant in a live feedback system—not the source of automation, but the interpreter of context. In doing so, it enables a shift from productivity tracking to co-regulatory alignment, reframing success as sustainable engagement rather than task throughput.
As human-computer interaction moves toward inclusive, context-aware systems, we argue that AI-based tools should not merely emulate neurotypical focus patterns, but accommodate neurodivergent rhythms. This reframing can help prevent AI from replicating normative biases in attention modeling and instead foster tools that uplift rather than pathologize cognitive diversity.

\subsection{Implications for Team Dynamics (future scope)}
While the current design focuses on individual support, the assistant’s underlying architecture and presence-based logic suggest possibilities for enhancing team-level awareness and coordination in neurodiverse teams. For example, a future iteration of the assistant could serve as a transparent layer of co-working presence, making it easier for teams to understand cognitive availability or engagement patterns without compromising privacy.

Features like shared “body doubling” blocks, reflective prompts during group planning or ambient status cues could help reduce friction in hybrid or asynchronous teams, especially when some members experience ADHD-related communication gaps, delays or social fatigue. These features must be designed with strict user agency and consent but they open up a rich future direction for neuroinclusive collaboration systems.

\subsection{Relevance for Wider Neurodivergent Support and Remote Work}
The assistant’s design principles—context sensitivity, privacy preservation, voice-based reflection and adaptive co-regulation—are relevant beyond ADHD-specific use cases. Individuals on the autism spectrum, those with anxiety or people experiencing executive dysfunction due to burnout or long-COVID may also benefit from systems that support rather than correct, accompany rather than monitor.

Furthermore, as remote and hybrid work become the norm, the challenges of digital distraction, emotional disconnection and unstructured work environments are becoming universal. The assistant's ability to create a quiet rhythm of presence and accountability without external surveillance offers a model for more humane digital work ecosystems—not only for neurodivergent individuals but for all users seeking stability in attention-fragmented environments.

\section{Limitations and Future work}

\subsection{Limitations}
While this paper introduces a novel, privacy-first cognitive assistant for ADHD-affected professionals, it is not without limitations. First, the system is currently conceptual and in a very early stage of prototyping, with no deployment data. While initial user preferences were captured via a 25-participant survey and author lived experience, more rigorous long-term user testing is required to evaluate real-world engagement, behavioral outcomes and emotional safety over time.

Second, the assistant’s on-device ML models are currently defined as rules-based and interpretable. Although suitable for privacy and explainability, more complex behaviors and patterns (e.g., emotional drift, motivational spirals) may benefit from personalized ML models, requiring federated learning or encrypted local training — which remains an open challenge.

While our co-design approach provided valuable qualitative insights, the broader spectrum of neurodivergent experiences is still not fully represented. We believe that incorporating perspectives from individuals with other conditions such as autism and from more diverse cultural backgrounds would make the assistant more inclusive and widely relevant.

\subsection{Future work}
As attention, autonomy and ambient presence become critical dimensions of digital life, the tools we design must move beyond automation and toward relational intelligence—co-regulating with us, not just computing for us.

Future work will focus on three promising directions:
\begin{itemize}
    \item Functional Prototyping and User Testing: A deployable prototype will be developed for controlled in-situ testing with ADHD-affected professionals. This will allow for real-time data collection on engagement patterns, the effectiveness of nudges and perceived emotional support. User journaling and A/B comparisons with static tools may provide deeper insight into assistant efficacy.
    \item Shared Body Doubling for Teams (Privacy-Preserving Multi-User Co-Regulation):
A promising future direction is extending the assistant for neuroinclusive teams, where multiple ADHD or neurodivergent users may opt into a shared digital presence—such as synchronized body doubling, co-check-ins or  group reflection prompts. Research on mutual body doubling (Eagle et al., 2024) \cite{eagle_proposing_2023} suggests that peer-based co-regulation is not only effective but emotionally meaningful. A privacy-first assistant could facilitate these interactions without exposing individual attention data, using ambient group cues or synchronized availability windows.
\item Cross-Domain Adaptation:
While this assistant is designed for professional digital work, its architecture is potentially applicable in adjacent domains such as remote education, creative workflows and digital therapy. Exploring its impact for users with long-COVID cognitive fatigue, anxiety or sensory regulation needs could broaden its role as a universal neuroadaptive co-regulator.
\end{itemize}

\section{Conclusion}
As digital work environments become increasingly fragmented and cognitively demanding, professionals affected by ADHD continue to navigate persistent challenges—task switching, time blindness, emotional dysregulation and executive dysfunction—often without meaningful support. Our survey findings echo what lived experience already suggests: that existing tools, while well-intentioned, tend to focus on behavioral correction, rigid task tracking or  notification-driven reminders, which can increase stress and lead to disengagement.

ADHD-affected IT workers, especially in remote or hybrid roles, often work in isolation. The stigma surrounding neurodivergence means many are hesitant to disclose their needs to colleagues or managers, making self-regulation both a private burden and an invisible barrier to productivity. Moreover, with increased digital exposure across both early-career and experienced professionals, the risk of attention fatigue and emotional burnout is not only growing—it is being normalized.

In response, this paper has proposed a privacy first, voice-enabled productivity assistant that reimagines support for ADHD-affected professionals. Rather than enforcing compliance, the assistant functions as a context-aware, soft-touch co-regulator, offering presence, reflection and adaptive nudging based on real-time behavioral signals all processed locally and under user control. Its inclusion of a digital body doubling mode, based on community-driven coping strategies, affirms that neuroinclusive design must move beyond features and toward empathetic experience modeling.

Ulimately, this assistant is not just a tool it is a proposition: that cognitive support technologies can be quiet, respectful, emotionally intelligent and aligned with lived neurodivergent realities. As attention becomes the new currency of digital work, we must ask not just how we manage it.

\bibliographystyle{ACM-Reference-Format}
\bibliography{ADHD-Paper-BibFile}

\end{document}